\documentclass[aps,twocolumn,pra,amsfonts,superscriptaddress,floatfix,10pt]{revtex4}
\expandafter\let\csname equation*\endcsname\relax
\expandafter\let\csname endequation*\endcsname\relax
\usepackage{amsmath}

\usepackage{bm}
\usepackage{float}
\usepackage{mathtools}
\usepackage{multirow}

\usepackage{bbold}
\usepackage{bbm}
\usepackage{braket}

\usepackage{amsthm}
\usepackage{mathrsfs}
\usepackage[usenames]{xcolor}
\usepackage{gensymb}
\usepackage{appendix}
\usepackage{graphicx}
\usepackage[caption=false]{subfig}
\usepackage[normalem]{ulem}
\usepackage{color}

\usepackage{soul}
\usepackage{hyperref}

\usepackage{cancel}

\def\beq{\begin{equation}}
\def\eeq{\end{equation}}
\def\bea{\begin{eqnarray}}
\def\eea{\end{eqnarray}}

\begin{document}

\title{Time Crystals in the Driven Transverse Field Ising Model under Quasiperiodic Modulation}


\author{Pengfei Liang}
\affiliation{Beijing Computational Science Research Center, Beijing, China}
\affiliation{Abdus Salam ICTP, Strada Costiera 11, I-34151 Trieste, Italy}

\author{Rosario Fazio}
\affiliation{Abdus Salam ICTP, Strada Costiera 11, I-34151 Trieste, Italy}
\affiliation{Dipartimento di Fisica, Universit\`a di Napoli ``Federico II'', Monte S. Angelo, I-80126 Napoli, Italy}
\thanks{On leave}
\affiliation{Beijing Computational Science Research Center, Beijing, China}

\author{Stefano Chesi}
\affiliation{Beijing Computational Science Research Center, Beijing, China}
\affiliation{Abdus Salam ICTP, Strada Costiera 11, I-34151 Trieste, Italy}

\begin{abstract}

We investigate the transverse field Ising model subject to a two-step periodic driving protocol and quasiperiodic modulation of the Ising couplings. Analytical results on the phase boundaries associated with Majorana edge modes and numerical results on the localization of single-particle excitations are presented.
The implication of a region with fully localized domain-wall-like excitations in the parameter space is eigenstate order and exact spectral pairing of Floquet eigenstates, based on which we conclude the existence of time crystals. We also examine various correlation functions of the time crystal phase numerically, in support of its existence.  

\end{abstract}

\maketitle

\section{Introduction}

Our understanding of the out of equilibrium phase structures of periodically driven (Floquet) quantum many-body systems has made impressive progress over the past decade, see
for example the review~\cite{bukov2015} and the references therein. Among the many interesting discoveries made in this context, the prediction and the consequent observation of 
a time-crystalline phase predicted by Wilczek~\cite{wilczek2012prl1,wilczek2012prl2,bruno2013prl2,nozieres2013el,bruno2013prl3,sacha2015pra} in Floquet systems has stimulated 
a lot of interest.  The first concrete proposal to circumvent the no-go theorem by Oshikawa and Watanabe~\cite{oshikawa2015prl}, assessing the impossibility to have spontaneous 
breaking in equilibrium systems, was to consider periodically driven systems. Floquet time-crystal were predicted theoretically in~\cite{nayak2016prl,sondhi2016prl} and soon 
after experimentally observed in~\cite{monroe2017nature,lukin2017nature}. In the last few years the literature on time-crystals has grown enormously in many different 
directions, see for example~\cite{yao2017,russomanno2017,else2017,pal2018,iemini2018,rovny2018,surace2019,wu2019,schaefer2019,choi2019,iadecola2019,buca2019,zhu2019,
seibold2020,pizzi2019,matus2019,Heyl2020}. A recent overview can be found in~Ref.~\cite{sacha2018}.

Unlike previous works on manipulating and engineering effective Hamiltonians in the prethermal stage of Floquet evolution building upon high-frequency expansion~\cite{eckardt2017rmp} 
and on the edge states in Floquet systems relating to non-trivial topology of bulk bands, the long-time state of the discrete time crystals (DTC) is characterized by persistent oscillation of 
local order parameter, which implies the existence of non-trivial bulk spatio-temperal order. On the eigenstate level, it manifests eigenstate order in that almost all Floquet eigenstates possess 
non-trivial long range order.

Generic non-integrable Floquet systems will be heated up to the infinite temperature ensemble due to persistent pumping of energy by the driving \cite{rigol2014prx,lazarides2014pre,
rosch2015pra}, while the long time behavior of integrable Floquet systems is described by the periodic Gibbs ensemble~\cite{russomanno2012prl,lazarides2014prl,sen2019prb}. This fact implies 
that the long time steady states of both cases always have trivial correlations, thus excludes any non-trivial phase structures. A way out, as first proposed in Ref.~\cite{nayak2016prl}, is to consider many-body localized (MBL) systems~\cite{basko2006annp,huse2007prb}. 
MBL in the presence of high-frequency periodic drive was considered as well~\cite{abanin2015prl,abanin2016aop}. Many-body eigenstates may be classified by their broken symmetries 
and topology in the MBL phase \cite{vishwanath2015naturec,huse2007prb,potter2016prb,else2016prb}, analogous to that of phases transitions in equilibrium systems. 
Following this line, Ref.~\cite{yao2017prl} showed that the driven quantum Ising chain hosts a $Z_2$ time crystal phase.

Localization can be induced not only by quenched disorder, but also by quasiperiodic modulation of couplings. As first discussed by Azbel~\cite{azbel1979prl}, Aubry and 
Andre~\cite{andre1980aips} and their generalizations \cite{thouless1983prb,wilkinson1984prsa,sokoloff1985physrep,kohmoto1989prl,igloi1993jpa},
incommensurability leads to a localization-delocalization transition even for the 1D tight-binding lattice model.
In this work, we study the out of equilibrium phase structure of a driven transverse field Ising model (TFIM) with quasiperiodic (QP) modulation of Ising coupling. We work 
with an integrable model which allows us to use various analytic tools and present reliable numerical results. Making use of the Jordan-Wigner (JW) transformation, the driven 
TFIM is mapped to a driven Majorana chain. We start by analyzing the Majorana edge modes and localization of single-particle excitations and show that localized domain-wall-like 
excitations lead to eigenstate order and time crystal order, therefore building a connection between localization of single-particle excitations and time crystals.

The paper is structured as follows. In Sec.~\ref{sec:model} we introduce our model and discuss the implication of its symmetries on the phase structure. In Sec.~\ref{sec:results} we present our results regarding the Majorana edge modes, localization of single-particle excitations, long-range order of excited states in order. Based on these we conclude the existence of time crystal phase and present numerical results supporting our statement. We summarize our results in Sec.~\ref{sec:summary}.

\section{Model}\label{sec:model}

The driven QP-TFIM we consider consists of a two-step drive protocol, which is described by the following time-periodic Hamiltonian with period $T=T_J+T_b$
\begin{eqnarray}\label{eq:hamiltonian}
H=
\begin{cases}
b \sum_j\limits \sigma_j^z, \quad\quad\quad\quad 0\le t < T_b \\
\sum_j\limits J_j \sigma_j^x\sigma_{j+1}^x, \quad T_b\le t<T_b+T_J.
\end{cases}
\end{eqnarray}
Here $\sigma_j^{x,y,z}$ are the Pauli operators acting on site $j$ and the Ising couplings are given by a smooth function, which we take of sinusoidal form
\bea
J_j=J+A_J\sin(2\pi Qja+\varphi_J).
\eea
Quasiperiodicity implies that the wavelength $1/Q$ is incommensurate to the lattice constant $a$, namely $Q$ is an irrational number  (we set $a=1$ hereinafter). To analyze such a system theoretically, one can start by approximating the modulation by a sequence of periodic functions, which retrieve translational invariance. In this work we choose $Q$ as the golden mean $Q=(\sqrt5+1)/2$ and approximate it by the consecutive ratios of Fibonacci numbers $q_{n+1}/q_n$, where the Fibonacci 
sequence is given by the recurrence relation $q_{n+1}=q_{n}+q_{n-1}$ and the initial conditions $q_1=1, q_2=1$.

The undriven version of this QP-TFIM, with the same form of Ising couplings $J_j$, was studied in great detail in Ref.~\cite{laumann2017prx}, where it was found that in the strong modulation regime $J<A_J$ new gapless phases with localized or multifractal excitations emerge. In a similar way, the constant Ising coupling $J$ of our driven model controls the density of weak couplings and, as we will see below, plays a crucial role in determining the localization properties of (\ref{eq:hamiltonian}). We have confirmed in our numerics that the phase $\varphi_J$ is irrelevant to magnetic order and localization properties of single-particle excitations away from the phase boundaries, similar to what was observed in the undriven case \cite{laumann2017prx,note1}.

\subsection{Symmetries of the Floquet operator}

Our interest will be in the long-time behavior of the driven system, so it suffices to consider the Floquet operator of (\ref{eq:hamiltonian}), which reads
\bea\label{eq:floquetunitary}
U_F = \exp\left[-i T_J\sum_j\limits J_j \sigma_j^x\sigma_{j+1}^x\right]\exp\left[-i T_bb\sum_j\limits \sigma_j^z\right]. 
\eea
There are several symmetries of this Floquet operator. The $Z_2$ Ising symmetry, denoted by the symmetry operator $P=\prod_j \sigma_j^z$, is inherited from the Hamiltonian (\ref{eq:hamiltonian}) and  satisfies  $PU_FP^\dagger=U_F$. Furthermore,  the Floquet operator $U_F$ simply changes sign under the shift $T_JJ\to T_JJ+\pi$ or $T_bb\to T_bb+\pi$, which can be compensated by shifting the quasienergy $\epsilon\to\epsilon+\pi$ while leaving the Floquet eigenstate unchanged. This implies that all the physical properties have a periodicity $\pi$ in both $T_JJ$ and $T_bb$. Finally, we consider the reflections $T_bb\to -T_bb$ and $T_JJ\to-T_JJ$. For the $T_bb$ reflection, $U_F$ preserves its form after applying local rotations to all the sites, such that $\sigma_j^z\to-\sigma_j^z$. Instead, the $T_JJ$ reflection can be combined with $\varphi_J\to\varphi_J+\pi$ and local rotations on the even (or odd) sites, such that $\sigma_j^x\to-\sigma_j^x$. Taking into account all these symmetries, we can restrict our discussion to the rectangular region $[0,\pi]\times[0,\pi]$ in the $T_bb$-$T_JJ$ plane.  Another simple observation is that we can assume $A_J> 0$, as a change of sign is equivalent to a phase shift  $\varphi_J\to\varphi_J+\pi$.

\subsection{Majorana representation}

To describe the time evolution in terms of noninteracting fermions, we introduce a pair of Majorana fermions for each spin-$1/2$ 
 \begin{eqnarray}
\gamma_{2j}=\left(\prod_{i<j}\sigma_i^z\right)\sigma_j^x,\quad \gamma_{2j+1}=\left(\prod_{i<j}\sigma_i^z\right)\sigma_j^y. 
\end{eqnarray}
The Majorana operators satisfy the anticommutation relation $\{\gamma_i,\gamma_j\}=2\delta_{ij}$. Then, the Floquet operator can be expressed as $U_F = U_1U_2$, where
\bea \label{U12_def}
U_1&=&\exp(-T_J\sum_j\limits J_j \gamma_{2j+1}\gamma_{2j+2}) \nonumber \\
U_2&=&\exp(-T_bb \sum_j\limits \gamma_{2j}\gamma_{2j+1}).
\eea
Note also that there is a boundary term if periodic boundary conditions are imposed on the spin chain. However, this makes no trouble in our case as we will consider a semi-infinite chain in the study of edge modes, while for sufficient long chains the
boundary term is marginal in detecting bulk properties. The Majorana representation is superior to the fermionic one in that it allows us to treat the problem of edge modes and localization of excitations on equal footing.

A crucial feature of the Floquet operator is that it consists of two unitaries acting on disconnected Majorana pairs. We will see that this brings further simplifications, allowing us to obtain analytic expressions for some of the phase boundaries. For the moment, we derive the eigenequations for the Fermionic eigenmodes, which are defined in terms of the Majorana operators as
\begin{eqnarray}\label{Gamma_s}
\Gamma_s=\sum_{j=0}^\infty \alpha_j\gamma_{2j} + \sum_{j=0}^\infty \beta_j \gamma_{2j+1},
\end{eqnarray}
and satisfy $U^\dagger_F \Gamma_s U_F=e^{-i\epsilon_s}\Gamma_s$. Since shift of the quasienergy $\epsilon_s \to\epsilon_s+2\pi$ results in the same eigenmode, we can restrict the values of $\epsilon_s$ to the first Brillouin zone $[-\pi,\pi]$. In general the $\Gamma_s$ ($\Gamma_s^\dag$) are fermionic annihilation (creation) operators, satisfying $\{ \Gamma_s, \Gamma_{s'}^\dag \} = \delta_{s,s'} $. Since  $\Gamma_s$, $\Gamma_{s'}^\dag$ are associated to opposite quasienergies, in the following we further restrict $\epsilon_s\in[0,\pi]$.  
The action of the two unitaries $U_{1,2}$ on a single Majorana fermion can be easily written down [see Eq.~(\ref{U12_action})] and, with $j$ in the bulk, yields
\bea\label{eq:bulk}
\alpha_j c_{j-1} c_b+\alpha_{j+1} s_j s_b+\beta_j c_j s_b-\beta_{j-1} s_{j-1} c_b = e^{-i\epsilon_s}\alpha_j, \nonumber \\
-\alpha_j c_{j-1} s_b+\alpha_{j+1} s_j c_b+\beta_j c_j c_b+\beta_{j-1} s_{j-1} s_b = e^{-i\epsilon_s} \beta_j,
\eea
where we defined $c_j=\cos 2T_JJ_j, s_j=\sin 2T_JJ_j$ and $c_b=\cos 2T_bb, s_b=\sin 2T_bb$.  

Finally, based on Eq.~(\ref{eq:bulk}) we discuss the extension of the weak modulation condition to the driven Floquet system. In the time-independent model, one requires that $J_j$ does not attain arbitrarily small values (for arbitary $j$), simply leading to the condition $J>A_J$. Here, instead, we notice that the equations of motion (\ref{eq:bulk}) are left unchanged by the following transformation: $J\to 2\pi/T_J-J, \phi_J\to\phi_j+\pi,\epsilon\to\epsilon+\pi$, while $\alpha_j\to (-1)^j \alpha_j$ and $\beta_j\to (-1)^j \beta_j$. Because of such mapping between $J$ and $J\to 2\pi/T_J-J$, we define the weak modulation regime as follows
\bea\label{weak_coupling}
A_J < J <\frac{2\pi}{T_J}-A_J.
\eea
It is readily seen that Eq.~(\ref{weak_coupling}) recovers the expected condition $J>A_J$ of the undriven model when taking the appropriate limit of $T \to 0$. On the other hand, Eq.~(\ref{weak_coupling}) can only be satisfied for $T_JA_J< \pi/4$.

\section{Main Results}\label{sec:results}

\subsection{Majorana edge modes}\label{sec:edgemode}

To study the edge modes, we consider a semi-infinite chain, where it is possible to search for $\Gamma_s$ in the form of Majorana operators. For Floquet systems, such nontrivial edge states can lie at zero quasienergy or at the edge of the Brillouin zone ($\epsilon_s =  \pi$), and we denote them as $\Gamma_0$ and $\Gamma_\pi$, respectively. To find analytic expression for the phase boundaries associated with $\Gamma_{0,\pi}$, in \ref{sec:appendixa} we write explicitly the relevant eigenproblems in terms of the $\alpha_j,\beta_j$ coefficients of Eq.~(\ref{Gamma_s}). In particular, we find  that $U^\dagger_F \Gamma_0U_F=\Gamma_0$ is equivalent to the recurrence relation
\bea
\bold{v}_j=M_j \bold{v}_{j-1},
\eea
where we introduce the $2\times1$ vectors $\bold{v}_j=(\beta_j,\alpha_{j+1})^T$ and the $2\times2$ transfer matrices $M_j$,
\bea \label{Mj_definition}
M_j=\frac{1}{s_b}
\begin{bmatrix}
s_{j-1} & c_b - c_{j-1} \\
\frac{s_{j-1}(c_b -c_j)}{s_j} & \frac{1+c_{j-1} c_j-c_b(c_{j-1}+c_j)}{s_j}
\end{bmatrix},
\eea
where $c_j,s_j$ and $c_b,s_b$ are defined after Eq.~(\ref{eq:bulk}). Meanwhile, the starting vector $\bold{v}_0$ can also be evaluated explicitly as
\begin{eqnarray}
\bold{v}_0= \alpha_0 \frac{1-\cos 2T_bb}{\sin 2T_bb}
\begin{bmatrix}
-1 \\
\frac{1+\cos 2T_JJ_0}{\sin 2T_JJ_0}
\end{bmatrix},
\end{eqnarray}
where $\alpha_0$ is fixed by the normalization of $\Gamma_0$ and is irrelevant for computing the phase boundaries. 

\begin{figure}
  \begin{center}
    \begin{tabular}{c}
      \hspace{0cm}\resizebox{80mm}{!}{\includegraphics{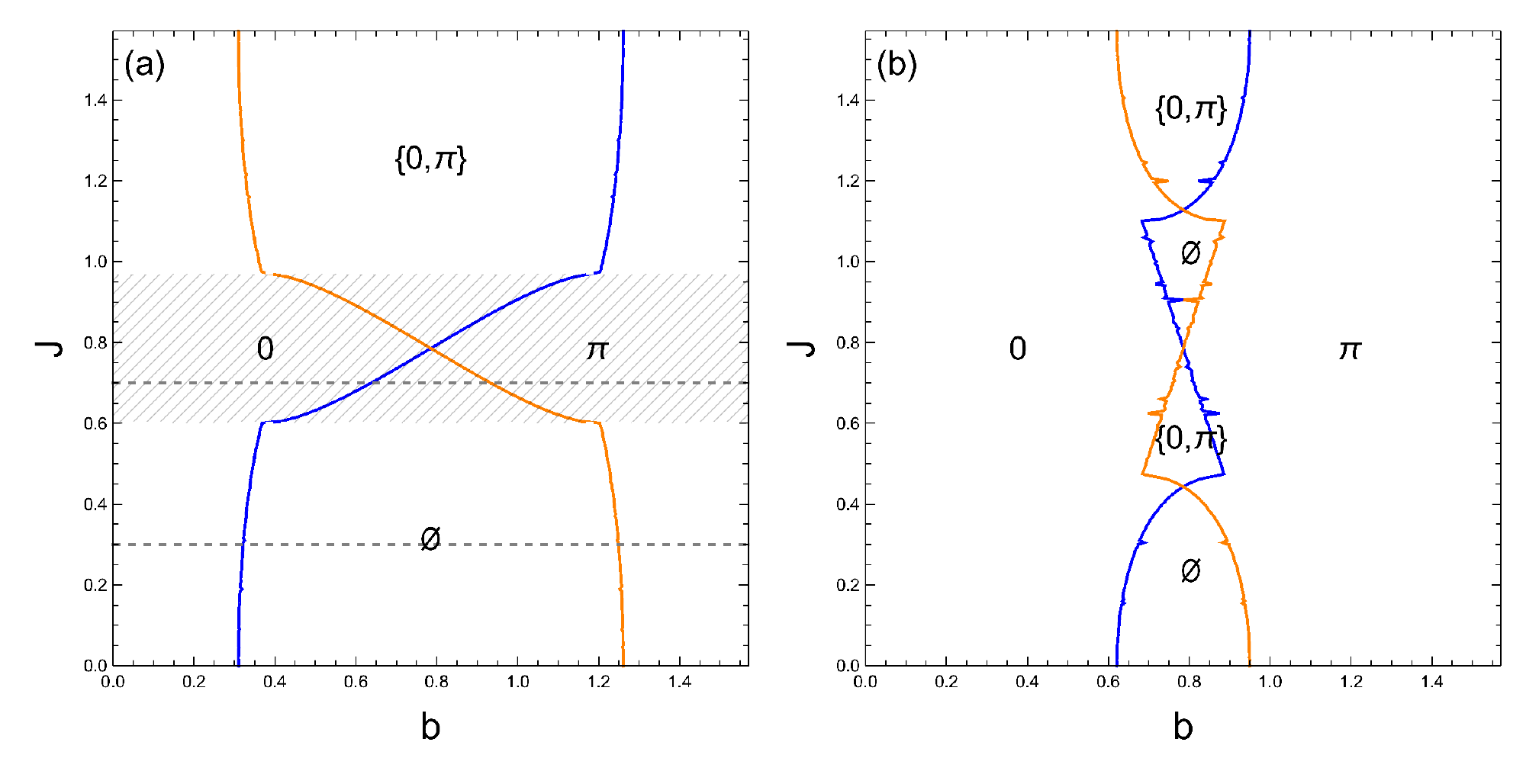}}\\
    \end{tabular}
  \end{center}
\caption{Phase diagrams for different $A_J$s. In (a) $A_J=0.6$ and in (b) $A_J=1.1$. Blue (Orange) solid lines indicates the phase boundary for $0$ $(\pi)$ edge mode and different phases are characterized by the number of edge modes ($\emptyset$ means that no edge mode exists). In (a), the parameter space is separated into weak modulation (hatched region) and strong modulation (white region) regimes, where the transitions fall into different universality classes. In (b), while the whole parameter space is strongly modulated,  we still observe two nonanalytic points on each phase boundary. Instead, the small spikes are due to limited numerical precision.}
\label{fig:pds}
\end{figure}

When $Q$ takes the value $q_{n+1}/q_n$, the transfer
matrix $M_j$ is periodic with respect to its index $j$ with period $q_n$, namely $M_j=M_{j+q_n}$. The total transfer matrix for one period is therefore
\begin{eqnarray}
\bold M_{q_n}= \prod_{j=1}^{q_n} M_j.
 \end{eqnarray}
In \ref{sec:appendixb} we prove that the vector $\bold{v}_0$ is an (unnormalized) eigenvector of the transfer matrix $\bold{M}_{q_n}$, and the corresponding eigenvalue $\lambda_{0,q_n}$ satisfies
\begin{equation}\label{eq:loglambda0}
\ln |\lambda_{0,q_n}|=q_n\left(\ln\left|\tan T_bb\right|+\frac{1}{q_n}\sum_{j=0}^{q_n-1}\ln\left|\cot T_JJ_j\right|\right).
\end{equation}
In the incommensurate limit $n\rightarrow\infty$, if $\lambda_{0}=\lim\limits_{n\rightarrow\infty}\lambda_{0,q_n}$ is less than $1$ there exists a Majorana eigenmode with zero quasienergy; otherwise, the edge mode does not exist. Therefore, the phase boundaries are found by setting the left-hand side of Eq.~(\ref{eq:loglambda0}) to zero. Furthermore, in the incommensurate limit the sum in the right hand side of Eq.~(\ref{eq:loglambda0}) can be approximated by an integral, giving
\begin{eqnarray}\label{eq:phaseboundary0}
-\ln\left|\tan b \right|=\int_0^{2\pi}\frac{d\theta}{2\pi}\ln\left|\cot J(\theta) \right|,
\end{eqnarray}
where $J(\theta)=J+A_J\cos\theta$ in the integrand and we set $T_b=T_J=1$, to simplify the notation (we follow this choice from now on). In the same manner, the equation to determine the phase boundary for the $\pi$ edge mode is 
\begin{eqnarray}\label{eq:phaseboundarypi}
-\ln\left|\cot b \right|=\int_0^{2\pi}\frac{d\theta}{2\pi}\ln\left|\cot J(\theta) \right|.
\end{eqnarray}
From Eqs.~(\ref{eq:phaseboundary0}) and (\ref{eq:phaseboundarypi}) we calculate the phase boundaries numerically and the phase diagram is depicted in Fig.~\ref{fig:pds} for different values of $A_J$. We observe that the parameter space is divided into distinct phases, characterized by the number of Majorana edge modes (or, equivalently, by different magnetic order of the vacuum state of fermionic excitations) 

In Fig.~\ref{fig:pds}(a) we have $A_J<\pi/4$ and the weak modulation regime [see Eq.~(\ref{weak_coupling})] is indicated by the hatched region. The phase boundaries determined by Eqs.~(\ref{eq:loglambda0}) and (\ref{eq:phaseboundarypi}) have a non-analytical behavior at the edges of the weak modulation region. This phase diagram coincides with that of the clean one if we take $A_J=0$, where the phase boundaries are two straight lines \cite{sondhi2016prl}. For weak modulation, all Ising couplings have the same sign and the phase transition is in the conventional Ising universality class, which is verified by the linear asymptotics of the spectrum around the transition points, see Fig.~\ref{fig:tbwIPR}(a). Thus, the dynamical exponent is $z=1$ in this case. On the other hand, strong modulation introduces a finite density of broken links and drives the transition unstable, making it fall into a new universality class. The numerically extracted exponent is $z\approx1.8$, see Fig.~\ref{fig:tbwIPR}(a). This value is close to that estimated in Refs.~\cite{laumann2018prl,laumann2018arxiv} using a saturated Harris-Luck criterion for the undriven TFIM, suggesting the two are in the same QP-Ising universality class.

In Fig.~\ref{fig:pds}(b) we show a representative phase diagram with $A_J>\pi/4$, when the Ising coupling are always strongly modulated and the phase boundaries are more complex. Here we still find non-analytic features marking the edges of a middle region in $J$, where we obtain a spectrum with vanishing gaps at both $0$ and $\pi$ quasienergies (not shown). On the other hand, Fig.~\ref{fig:tbwIPR}(a) shows that the spectrum in the weakly modulated region has gaps both around $0$ and $\pi$, and only one of the two gaps can disappear at the phase boundaries or in the strongly modulated region. The latter behavior is also found in the outer region of Fig.~\ref{fig:pds}(b). Based on this observation, we interpret the middle region of Fig.~\ref{fig:pds}(b) as overwhelmingly modulated. For our purpose of seeking time crystals, we will mainly focus on the case $A_J<\pi/4$ in the rest of the paper.

\subsection{Localization of Excitations}\label{sec:AL}

\begin{figure}
  \begin{center}
    \begin{tabular}{cc}
      \hspace{0.cm}\resizebox{80mm}{!}{\includegraphics{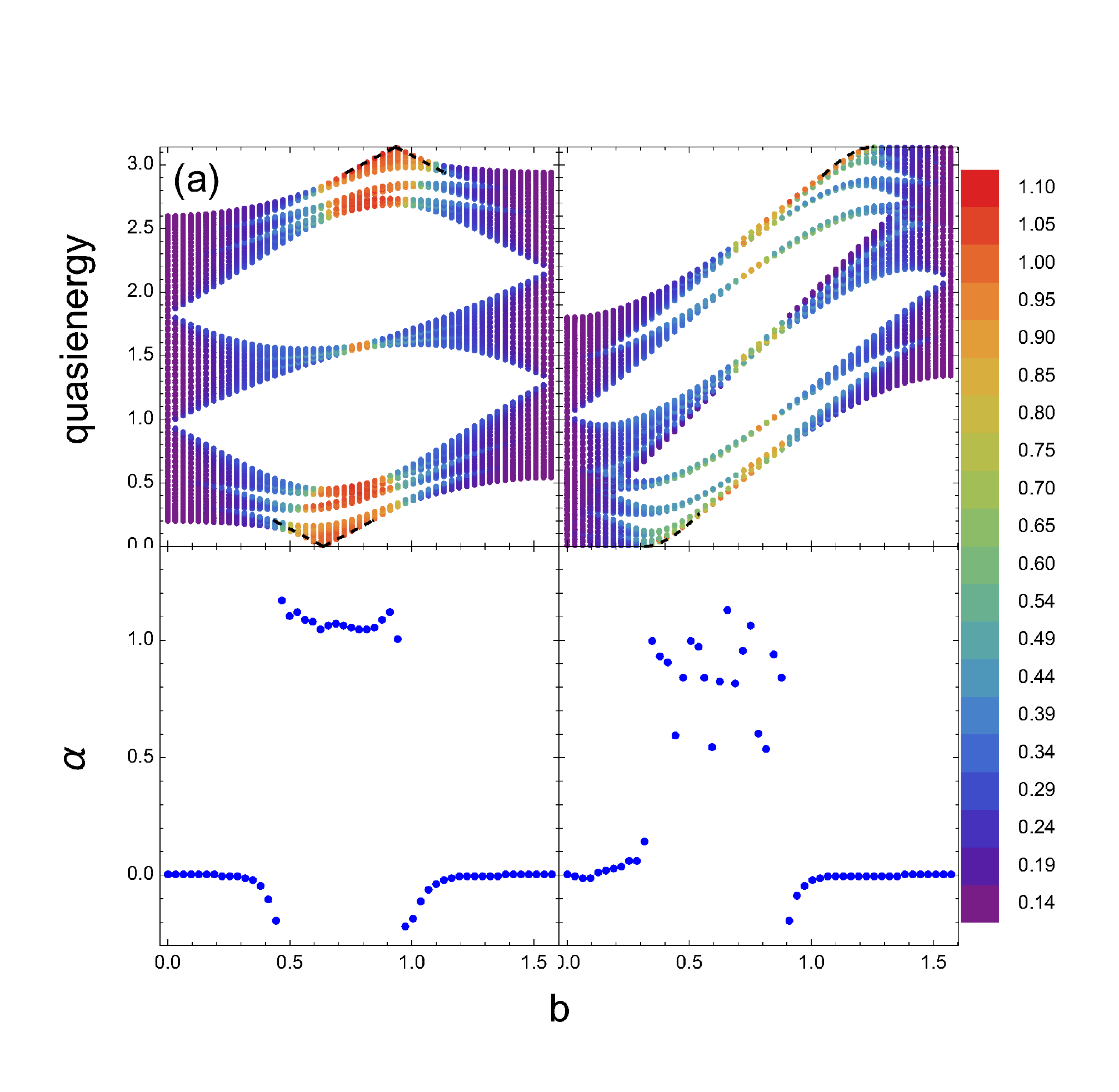}}\\
      \hspace{-0.cm}\resizebox{80mm}{!}{\includegraphics{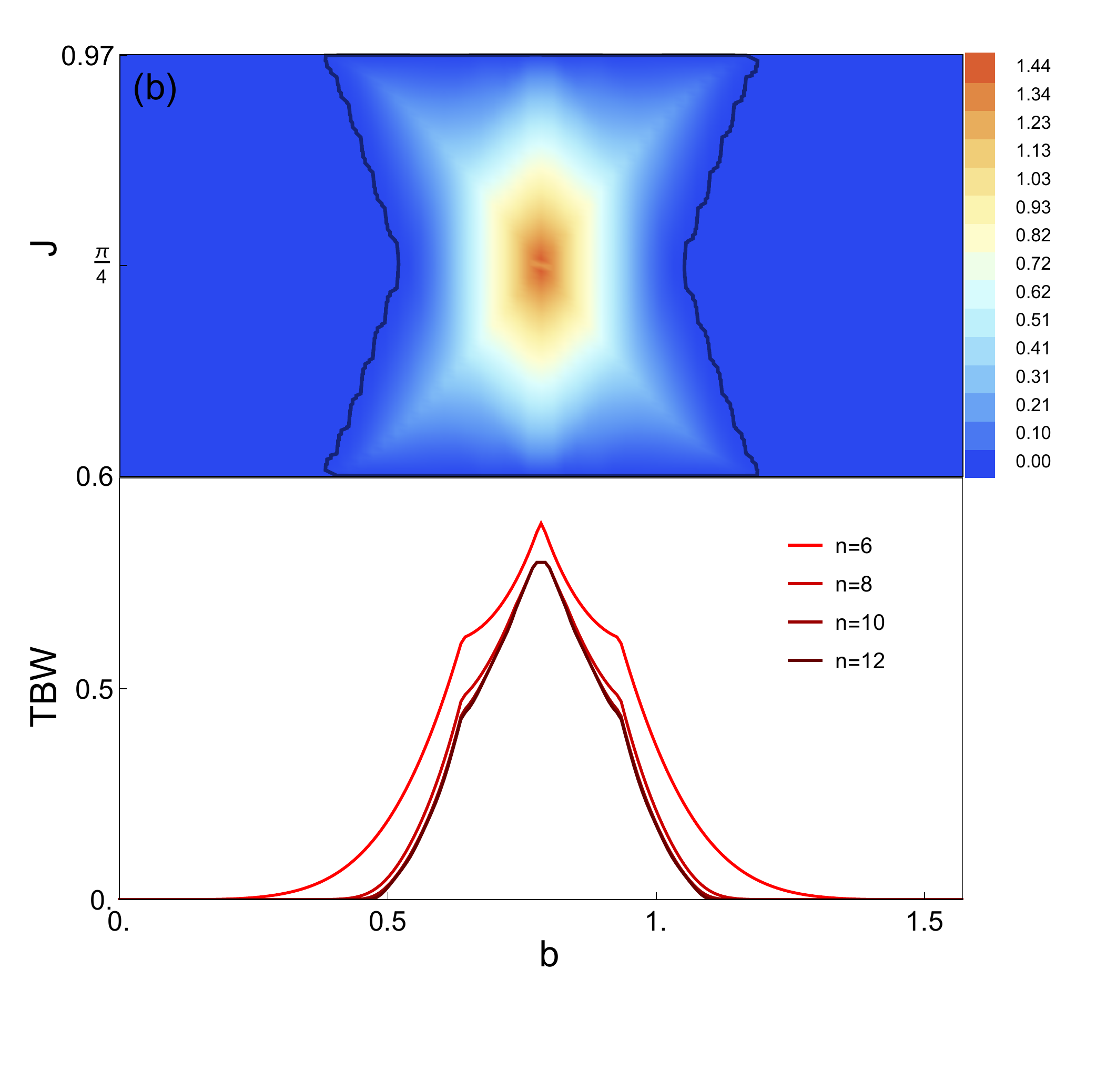}}
    \end{tabular}
  \end{center}
\caption{(a) Spectra, IPR of individual eigenstate, and its finite-size scaling for different values of $J$ (along the two line cuts in Fig.~\ref{fig:pds}(a)). In the left two panels $J=0.7$ in the weak modulation regime, while in the right two $J=0.3$ in the strong modulation regime. Color in the upper panels is given by the ratio -$\log\textrm{IPR}/\log q_n$, a rough estimate of the finite-size scaling exponent $\alpha$. Dashed lines mark how gaps close around transition points, from which we extract the dynamical exponent $z$. (b) TBW in the weak modulation regime for $A_J=0.6$ and a chain of length $q_n=144$ ($n=12$) with periodic boundary condition. Black solid lines in the upper panel are the numerically evaluated boundary of the two fully-localized phases. In the lower panel, we take a line cut at $J=0.7$ and show convergence of the TBW as a function of the linear size $q_n$.}
\label{fig:tbwIPR}
\end{figure}

The fact that the quasiperiodic Ising coupling can be approximated by a sequence of periodic potentials allows to impose an analytical upper bound for the spectral measure and to calculate numerically the sum of the bandwidth of all bulk bands in a finite chain, which we call the total bandwidth (TBW) here. The TBW essentially measures the fraction of extended states in the whole Floquet spectrum quantitatively. The analytical estimation in Ref.~\cite{laumann2018arxiv}, which is also applicatable to our case, shows that in the strong modulation regime (white region for $A_J<\pi/4$ and the whole parameter space for $A_J>\pi/4$) the spectral measure vanishes for infinite system size, excluding the existence of extended states. In the weak modulation regime, however, we refer to numerics to identify the localization-delocalization transitions (marked by black lines in Fig.~\ref{fig:tbwIPR}(b)). As expected, the states close to the lines of vanishing gap are forced to be extended, as a result of the Ising universality class, whereas in the vicinity of $b=0,\pi/2$ all states remain localized (marked by black lines in Fig.~\ref{fig:tbwIPR}(b)).

Another commonly adopted diagnostic of localization is the inverse participation ratio (IPR) calculated at different quasienergies, which gives an energy-resolved characterization of localization and is capable of detecting mobility edge. For a given Floquet eigenstate expressed in the Majorana representation, the IPR is defined by
\begin{eqnarray}
\textrm{IPR}=\sum_j\left(|\alpha_j|^4+|\beta_j|^4\right)
\end{eqnarray}
where $\alpha_j$ ($\beta_j$) is the amplitude on even (odd) Majorana site. Finite size scaling of the IPR at different energy density, $\textrm{IPR}\sim1/q^\alpha$ with $q$ system size, provides an efficient method for detecting localization. For extended states, $\alpha=1$; for localized states $\alpha=0$; while for critical states this exponent lies in between $0<\alpha<1$. We present numerical results in Fig.~\ref{fig:tbwIPR}(a) for the weak modulation regime (left two panels where $J=0.7$) and the strong modulation regime (right two panels where $J=0.3$) regarding the energy-resolved IPR and the exponent $\alpha$ of its finite size scaling relation. We see that in both cases the localization-delocalization transitions depend on energy, signaling the presence of a mobility edge. However, delocalized states in the two regimes are clearly different in that strong modulation brings all delocalized states into critical states with multifractional wavefunctions, coinciding with the theoretical prediction of vanishing spectral measure in this region.

\subsection{Long Range Magnetic Order of Excitations}\label{sec:LRO}

The phase diagram shown in Fig.~\ref{fig:pds}(a) can also be understood in terms of magnetic order of the vacuum state, satisfying $\Gamma_s|0\rangle = 0$. By conventionally choosing $\epsilon_s \in [0,\pi]$, the $\Gamma_s$ operators and their vacuum are uniquely defined. For simplicity, although the notion of ground state is meaningless for Floquet systems, we will sometimes refer to the eigenstates generated by $\Gamma_s^\dag$ as single-particle `excitations'. However, we emphasize that the choice of $|0\rangle$ has no special role in the present driven model. Instead, it is simply a reference state from which one can obtain the other Floquet eigenstates by applying the quasiparticle creation operators, and all the physical properties of the model are independent of the choice for $|0\rangle$.

\begin{figure}[h!]
  \begin{center}
    \begin{tabular}{c}
      \hspace{-0.cm}\resizebox{80mm}{!}{\includegraphics{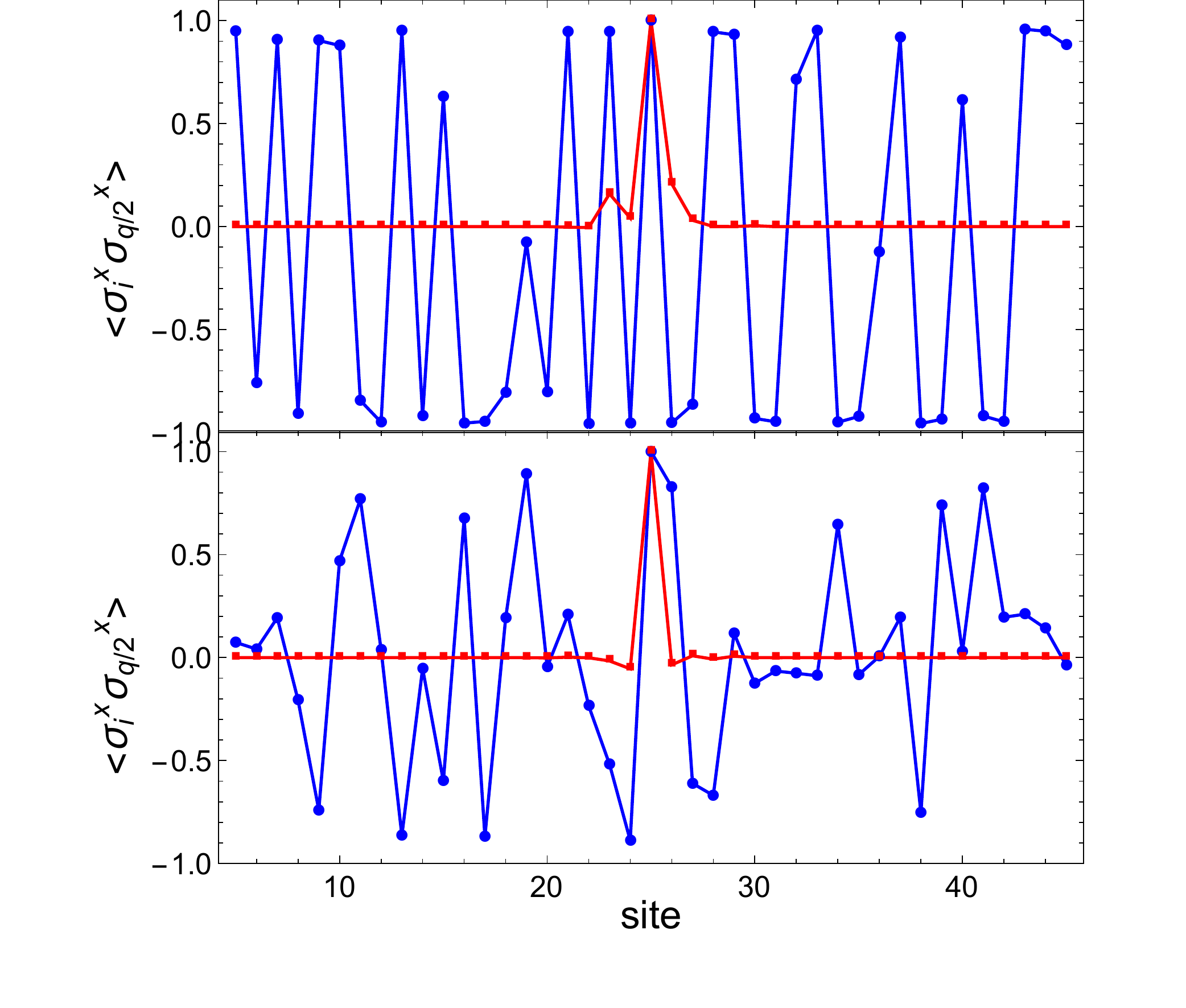}}
    \end{tabular}
  \end{center}

\caption{Correlation functions $\langle\sigma_i^x\sigma_j^x\rangle$ of a Floquet eigenstate with randomly chosen occupation numbers of the eigenmodes (blue curves). The upper and lower panels refer to the weak ($J=0.7$ and $A_J=0.6$) and strong ($J=0.3$ and $A_J=0.6$) modulation regimes, respectively. For comparison, we also plot the correlation function for the clean model
(red curves, with $A_J = 0$). Other parameters are: $b = 1.3$ and $\varphi_J = 0.4$.}
\label{fig:correlation}
\end{figure}

As in the case of the undriven TFIM \cite{sen1996}, no edge states imply trivial band topology and absence of long-range order, namely a paramagnetic (PM) phase; While one edge state, no matter if it is at $0$ or $\pi$ quasienergy, indicates nontrivial band topology and long-range ferromagnetic (FM) order. Finally, as discussed for the clean quantum Ising chain \cite{sondhi2016prb} ($J_j=J$), the new phase with two edge states, which is unique to Floquet systems, also has trivial bulk bands thus no long-range correlation in the bulk. Alternatively, in terms of spin observables, we can characterized different phases by the nature of their excited states. As we will discuss shortly below (see Fig.~\ref{fig:correlation}), the $``0"$ and $``\pi"$ phases eigenstates are composed of domain walls. Furthermore, due to the $Z_2$ symmetry (see the discussion in Sec.~\ref{sec:tc}), the eigenstates form pairs with fixed quasienergy splitting (zero and $\pi$, respectively for the two phases). For this reason, we shall also call these phases FM phases. On the other hand, domain walls are absent in the other two phases so we shall call them PM phases.

A remarkable consequence of the localization of excitations in the FM phases is the preservation of long-range order in a generic ``excited states", which is expressed as the Slater determinant of single-particle excitations. Nonvanishing long-range order in excited states follows from the fact that in the FM phase excitations are domain-wall-like, and a single domain wall simply revert the sign of the correlator of the vacuum state. In the localized phase, each excitation is composed of a few domain walls and is immobile. This explains what we see in Fig.~\ref{fig:correlation} where the correlation function changes sign many times, corresponding to the presence of many pinned domain walls. However, the correlation function never decay to zero at long distance. Long-range order protected by localization in almost all eigenstates is crucial in view of time crystal, as it helps to build the robust oscillation of magnetization in the dynamical process.

\subsection{$Z_2$ Time crystal Order}\label{sec:tc}

The second consequence of localized excitations is the realization of time crystals. First, let us point out  
the major difference between the two fully localized FM phases in the vicinity of $b=0$ and $\pi/2$. This is best illustrated in an open chain. Let us denote bulk Floquet excitations by $\Gamma_s$, in ascending order of the quasienergy $\epsilon_s$. Then any excited Floquet eigenstate is given by $|s_1,s_2,\cdots,s_j\rangle=\Gamma_{s_1}^\dagger\Gamma_{s_2}^\dagger\cdots\Gamma_{s_j}^\dagger|0\rangle$ hosting many excitations with $|0\rangle$ the aforementioned vacuum states. In the fully localized phase close to $b=0$ (see Fig.~\ref{fig:tbwIPR}), excited states come in nearly degenerate pairs, e.g., the pair $\Gamma_{s_1}^\dagger\Gamma_{s_2}^\dagger\cdots\Gamma_{s_j}^\dagger|0\rangle$ and $\Gamma_0^\dagger\Gamma_{s_1}^\dagger\Gamma_{s_2}^\dagger\cdots\Gamma_{s_j}^\dagger|0\rangle$;
On the contrary, in the fully localized phase close to $b=\pi/2$, excited states form pairs with $\pi$ splitting, namely $\Gamma_{s_1}^\dagger\Gamma_{s_2}^\dagger\cdots\Gamma_{s_j}^\dagger|0\rangle$ and $\Gamma_\pi^\dagger\Gamma_{s_1}^\dagger\Gamma_{s_2}^\dagger\cdots\Gamma_{s_j}^\dagger|0\rangle$. This spectral pairing becomes exact in the thermodynamic limit (TDL). Remarkably, spectral pairing with a $\pi$ quasienergy splitting is a prerequisite for the formation of $Z_2$ time crystals, where the discrete time translation symmetry is spontaneously broken and period doubling of the order parameter is observed. Combined with the long-range FM order in any Floquet excited states, we may anticipate that the fully localized FM phase close to $b=\pi/2$ is a $Z_2$ time crystal.

\begin{figure}[h!]
  \begin{center}    \begin{tabular}{ccc}
      \hspace{-0.cm}\resizebox{80mm}{!}{\includegraphics{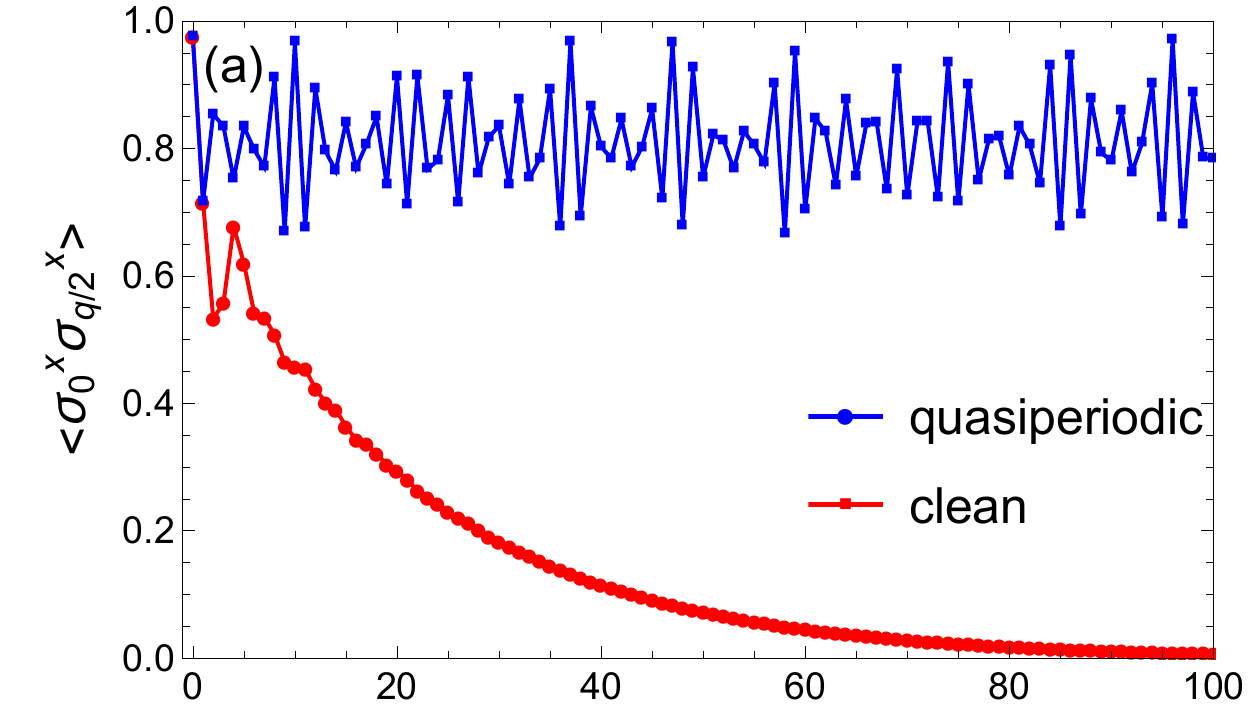}}\\
      \hspace{0.cm}\resizebox{80mm}{!}{\includegraphics{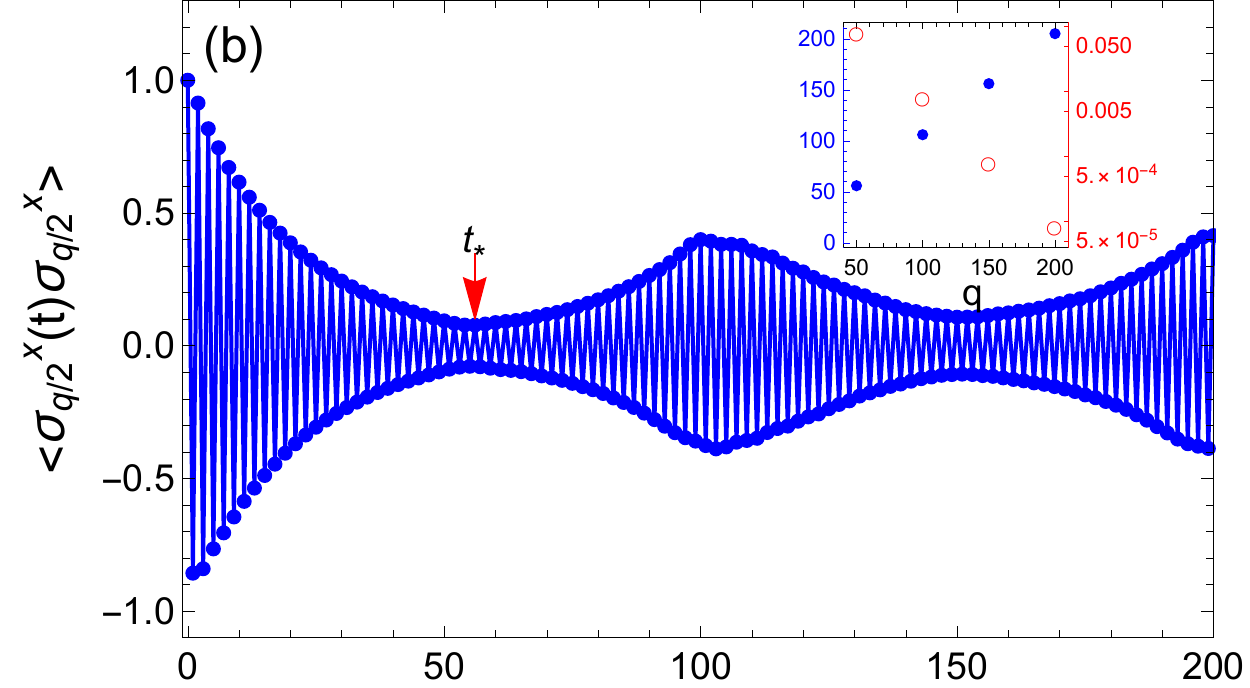}}\\
      \hspace{0.cm}\resizebox{80mm}{!}{\includegraphics{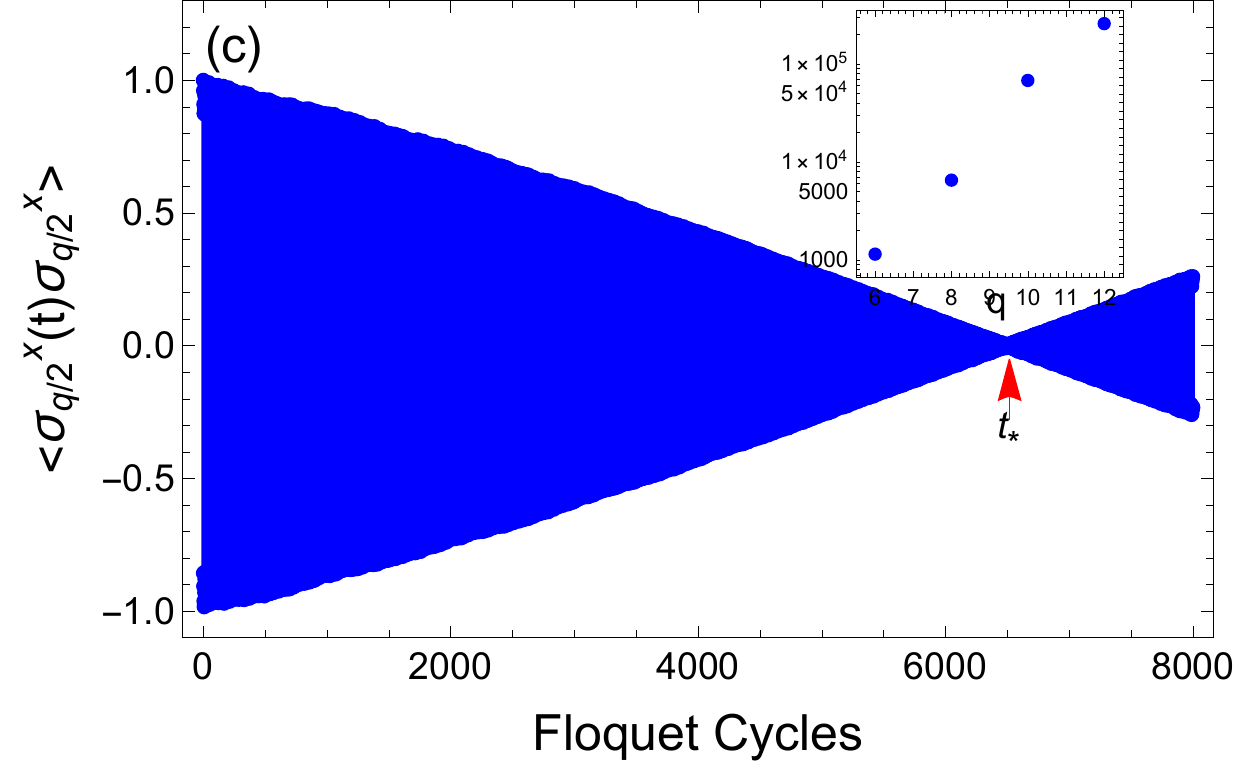}}\\
    \end{tabular}
  \end{center}

\caption{(a) Dynamics of the end-center equal-time correlation function for the clean (red) and quasiperiodic (blue) TFIMs with length $q=100$. (b) and (c) shows autocorrelation functions of the central spin for the clean and quasiperiodic cases respectively. To display the finite-size effect clearly, we choose a chain of lengths $q=50$ in (b) and $q=8$ in (c). Open boundary condition is adopted in all plots. Insets show the finite size scaling of the lifetime $t_*$ (blue dots) and that of the value of autocorrelation function at the life time $t_*$ (red circle).}
\label{fig:dyncorr}
\end{figure}

We can establish the $Z_2$ time crystal order by examining the dynamics of correlation functions, which are amenable to numerical calculations as they respect the Ising symmetry.
The equal-time correlation $\langle\psi(nT)|\sigma_i^x\sigma_j^x|\psi(nT)\rangle$ detects the presence or absence of FM order in the long run. In Fig.~\ref{fig:dyncorr} we plot the correlator $\langle\psi(nT)|\sigma_i^x\sigma_j^x|\psi(nT)\rangle$ as a function of time starting from the symmetry-unbroken ground state of a clean TFIM, for both the clean and the quasiperiodic modulated cases. For the clean model, the correlator decays exponentially to zero, indicating vanishing long-range order in the stationary state; while for the quasiperiodic model, it quickly relaxes to a nonvanishing value with some persistent fluctuation. In fact, expanding the initial state in the basis of Floquet eigenstates $|\psi(0)\rangle=\sum_{\alpha,s} C_{\alpha,s}|\alpha,s\rangle$ where the index $s$ accounts for possible degeneracy of eigenstates, we can separate the stationary value from the oscillation as \cite{santoro2012prl}
\begin{eqnarray}
\langle\sigma_i^x\sigma_j^x\rangle(nT)=\langle\sigma_i^x\sigma_j^x\rangle_D+\int d\Omega e^{-in\Omega T}f(\Omega),
\end{eqnarray}
where $\langle\sigma_i^x\sigma_j^x\rangle_D=\sum_{\alpha,s,s'} C_{\alpha,s'}^*C_{\alpha,s}\langle\alpha,s'|\sigma_i^x\sigma_j^x|\alpha,s\rangle$ denotes the generalized ``diagonal ensemble" when the systematic degeneracy of Floquet eigenstates $|\alpha,s\rangle$ is taken into consideration, and the function $f(\Omega)=\sum_{\alpha'\neq\alpha,s',s} e^{i(E_{\alpha',s'}-E_{\alpha,s})t}C_{\alpha',s'}^*C_{\alpha,s}\langle\alpha',s'|\sigma_i^x\sigma_j^x|\alpha,s\rangle\delta(\Omega-E_{\alpha',s'}+E_{\alpha,s})$
is the density of states weighted by the amplitudes of the initial state. For integrable systems (like models solvable by the JW transformation), the fluctuation in the equal-time correlation function is solely determined by $f(\Omega)$ and is related to the property of the single-particle spectrum \cite{santoro2012prl,santoro2013prb}. A continuous single-particle spectrum implies vanishing fluctuation in the long-time limit whereas a pure-point spectrum implies persistent fluctuation even in the TDL. Thus we conclude that for the quasiperiodic chain the equal-time correlator fluctuates around the diagonal ensemble value $\langle\sigma_i^x\sigma_j^x\rangle_D$ with no decay for any system size. We have verified this statement is true (not shown here) for very long time. 

The equal-time correlation function cannot distinguish the time crystal phase from a normal FM phase. This can be remedied by examining the autocorrelation function $\langle\sigma_j^x(nT)\sigma_j^x\rangle$. In Fig.~\ref{fig:dyncorr}(b) and (c), we plot the autocorrelation function of the central spin for the clean and quasipeioridic chains. In both cases, we see the autocorrelation function decays at an early stage then displays revival due to finite size effect. We define the time $t_*$ at which the autocorrelation function decays to its minimum for the first time as the lifetime of the time crystal in a finite system and investigate its scaling with the system size (see insets in Fig.~\ref{fig:dyncorr}). For the clean case, the lifetime scales linearly with the size as $t_*\sim q$ while for the quasiperiodic case it scales exponentially as $\ln t_*\sim q$. The exponential dependence of $t_*$ on system size $q$ supports a stable time crystal phase in the TDL.

\section{Conclusions}\label{sec:summary}

We have investigated the integrable TFIM subject to periodic driving and quasiperiodic modulation of the Ising coupling. Thanks to the JW transformation and the special structure of the Floquet operator, we were able to obtain analytical expressions for the Majorana edge modes. By applying the arguments in Ref.~\cite{laumann2018arxiv}, we show that in the strong modulation regime the spectral measure vanishes everywhere in the parameter space, while we resort to numerics in the investigation of spectral measure in the weak modulation regime and show that two fully-localized phases exist close to the two special values $b=0,\pi$. We analyzed two consequences of fully-localized single-particle excitations, namely long-range FM order of eigenstates and $\pi$ spectral pairing, based on which a $Z_2$ time crystal phase was anticipated. We also presented numerical results for relatively large systems in support of the existence of time crystals in our model. Our work will be a good starting point for future works on robustness of the time-crystalline order when integrability-breaking perturbations are included.

\section*{Acknowledgments}
R. F. acknowledges partial financial support from the Google Quantum Research Award. R. F. research has been conducted within the framework of the Trieste Institute for
Theoretical Quantum Technologies (TQT). S.C. acknowledges support from the National Key Research and Development Program of China (Grant No. 2016YFA0301200), NSFC (Grants No. 11974040 and No. 1171101295), and NSAF (Grant No. U1930402).

\appendix

\section{Transfer Matrix for Edge modes}\label{sec:appendixa}

We show here how to get the transfer matrix $M_j$ from the eigenequations for the  edge modes of a semi-infinite chain. As the $U_1$ and $U_2$ [defined in Eq.~(\ref{U12_def})] consist of odd-even and even-odd Majorana pairs respectively, we can write down explicitly their action on a single Majorana fermion,
\bea \label{U12_action}
U_1^\dagger \gamma_{2j}U_1&=&\gamma_{2j}\cos 2T_JJ_{j-1}+\gamma_{2j-1}\sin 2T_JJ_{j-1} \nonumber\\
U_1^\dagger \gamma_{2j-1}U_1&=&\gamma_{2j}\cos 2T_JJ_{j-1}-\gamma_{2j-1}\sin 2T_JJ_{j-1} \nonumber\\
U_2^\dagger \gamma_{2j}U_2&=&\gamma_{2j}\cos 2T_bb-\gamma_{2j+1}\sin 2T_bb \nonumber\\
U_2^\dagger \gamma_{2j+1}U_2&=&\gamma_{2j}\cos 2T_bb+\gamma_{2j+1}\sin 2T_bb. 
\eea
With these, one can easily derive the eigenequations for the $\alpha_j,\beta_j$. When $U_F^\dagger\Gamma_{0(\pi)} U_F=\pm\Gamma_{0(\pi)}$, Eq.~(\ref{eq:bulk}) for sites in the bulk $j\ge1$ simply becomes
\begin{align}\label{eq:eeMj}
\alpha_j c_{j-1} c_b+\alpha_{j+1} s_j s_b+\beta_j c_j s_b-\beta_{j-1} s_{j-1} c_b = \pm\alpha_j, \nonumber \\
-\alpha_j c_{j-1} s_b+\alpha_{j+1} s_j c_b+\beta_j c_j c_b+\beta_{j-1} s_{j-1} s_b = \pm\beta_j.
\end{align}
For $j=0$ the above equations should be replaced by
\bea\label{eq:eeedge}
\begin{aligned}
\alpha_0 c_b+\alpha_{1} s_0 s_b+\beta_0 c_{0} s_b= \alpha_0 , \\
-\alpha_0 s_b+\alpha_{1} s_{0} c_b+\beta_0 c_{0} c_b = \beta_0.
\end{aligned}
\eea
Eqs.~(\ref{eq:eeMj}) gives the transfer matrix $M_j$ in the main text, whereas solving Eqs.~(\ref{eq:eeedge}) leads to the solution of the starting vector $\mathbf{v}_0$.

\section{Eigenvalues and Eigenvectors of the Transfer Matrix $\bold{M}_{q_n}$}\label{sec:appendixb}

Here we prove that the vector $\bold{v}_0$ is an eigenvector of the transfer matrix $\bold{M}_{q_n}$, with its eigenvalue satisfying Eq.~(\ref{eq:loglambda0}) of the main text. By dropping an irrelevant prefactor, the vector $\bold{v}_0=(-s_0,1+c_0)^T$ can be written as a special case of
\begin{equation}
\bold{v}_j=
\begin{bmatrix}
-s_j \\ 1+c_j 
\end{bmatrix}.
\end{equation}
By direct calculation, it is easy to check that
\bea
M_{j+1}\bold{v}_{j}=\frac{1}{s_bs_{j+1}}(1-c_b)(1+c_{j})\bold{v}_{j+1}.
\eea
Therefore, after multiplying $\bold{v}_0$ by all of $M_j$s consecutively, we arrive at
\bea
\bold{M}_{q_n}\bold{v}_0&=&\left(\frac{1-c_b}{s_b}\right)^{q_n}\prod_{j=1}^{q_n}\frac{1-c_{j-1}}{s_j}
\bold{v}_{q_n} \nonumber\\
&=& \left(\frac{1-c_b}{s_b}\right)^{q_n}\prod_{j=0}^{q_n-1}\frac{1-c_j}{s_j}
\bold{v}_0,
\eea
where we made use of the periodicity of the Ising coupling $J_j=J_{j+q_n}$ in the second line. The eigenvalue can be written explicitly as
\begin{align}\label{eq:lambda0}
\lambda_{0,q_n}=\left(\tan T_bb\right)^{q_n}\prod_{j=0}^{q_n-1}\cot T_JJ_j,
\end{align}
and taking the absolute value and logarithm on both sides yields Eq.~(\ref{eq:loglambda0}). Note that, due to $\text{det}\;\bold{M}_{q_n}=1$ the other eigenvalue should be $1/\lambda_{0,q_n}$ and the corresponding eigenvector can be obtained readily.

\bibliography{driven_tfim_refs}

\providecommand{\newblock}{}
\begin{thebibliography}{10}
\expandafter\ifx\csname url\endcsname\relax
  \def\url#1{{\tt #1}}\fi
\expandafter\ifx\csname urlprefix\endcsname\relax\def\urlprefix{URL }\fi
\providecommand{\eprint}[2][]{\url{#2}}

\bibitem{bukov2015}
Bukov M, D'Alessio L and Polkovnikov A 2015 {\em Advances in Physics\/} {\bf
  64} 139--226 (\textit{Preprint}
  \eprint{https://doi.org/10.1080/00018732.2015.1055918})
  \urlprefix\url{https://doi.org/10.1080/00018732.2015.1055918}

\bibitem{wilczek2012prl1}
Shapere A and Wilczek F 2012 {\em Phys. Rev. Lett.\/} {\bf 109}(16) 160402
  \urlprefix\url{https://link.aps.org/doi/10.1103/PhysRevLett.109.160402}

\bibitem{wilczek2012prl2}
Wilczek F 2012 {\em Phys. Rev. Lett.\/} {\bf 109}(16) 160401
  \urlprefix\url{https://link.aps.org/doi/10.1103/PhysRevLett.109.160401}

\bibitem{bruno2013prl2}
Wilczek F 2012 {\em Phys. Rev. Lett.\/} {\bf 109}(16) 160401
  \urlprefix\url{https://link.aps.org/doi/10.1103/PhysRevLett.109.160401}

\bibitem{nozieres2013el}
Nozi{\`{e}}res P 2013 {\em {EPL} (Europhysics Letters)\/} {\bf 103} 57008
  \urlprefix\url{https://doi.org/10.1209%2F0295-5075%2F103%2F57008}

\bibitem{bruno2013prl3}
Bruno P 2013 {\em Phys. Rev. Lett.\/} {\bf 111}(7) 070402
  \urlprefix\url{https://link.aps.org/doi/10.1103/PhysRevLett.111.070402}

\bibitem{sacha2015pra}
Sacha K 2015 {\em Phys. Rev. A\/} {\bf 91}(3) 033617
  \urlprefix\url{https://link.aps.org/doi/10.1103/PhysRevA.91.033617}

\bibitem{oshikawa2015prl}
Watanabe H and Oshikawa M 2015 {\em Phys. Rev. Lett.\/} {\bf 114}(25) 251603
  \urlprefix\url{https://link.aps.org/doi/10.1103/PhysRevLett.114.251603}

\bibitem{nayak2016prl}
Else D~V, Bauer B and Nayak C 2016 {\em Phys. Rev. Lett.\/} {\bf 117}(9) 090402
  \urlprefix\url{https://link.aps.org/doi/10.1103/PhysRevLett.117.090402}

\bibitem{sondhi2016prl}
Khemani V, Lazarides A, Moessner R and Sondhi S~L 2016 {\em Phys. Rev. Lett.\/}
  {\bf 116}(25) 250401
  \urlprefix\url{https://link.aps.org/doi/10.1103/PhysRevLett.116.250401}

\bibitem{monroe2017nature}
Zhang J, Hess P~W, Kyprianidis A, Becker P, Lee A, Smith J, Pagano G,
  Potirniche I~D, Potter A~C, Vishwanath A, Yao N~Y and Monroe C 2017 {\em
  Nature\/} {\bf 543} 217--220 ISSN 1476-4687
  \urlprefix\url{https://doi.org/10.1038/nature21413}

\bibitem{lukin2017nature}
Choi S, Choi J, Landig R, Kucsko G, Zhou H, Isoya J, Jelezko F, Onoda S, Sumiya
  H, Khemani V, von Keyserlingk C, Yao N~Y, Demler E and Lukin M~D 2017 {\em
  Nature\/} {\bf 543} 221--225 ISSN 1476-4687
  \urlprefix\url{https://doi.org/10.1038/nature21426}

\bibitem{yao2017}
Yao N~Y, Potter A~C, Potirniche I~D and Vishwanath A 2017 {\em Phys. Rev.
  Lett.\/} {\bf 118}(3) 030401
  \urlprefix\url{https://link.aps.org/doi/10.1103/PhysRevLett.118.030401}

\bibitem{russomanno2017}
Russomanno A, Iemini F, Dalmonte M and Fazio R 2017 {\em Phys. Rev. B\/} {\bf
  95}(21) 214307
  \urlprefix\url{https://link.aps.org/doi/10.1103/PhysRevB.95.214307}

\bibitem{else2017}
Else D~V, Bauer B and Nayak C 2017 {\em Phys. Rev. X\/} {\bf 7}(1) 011026
  \urlprefix\url{https://link.aps.org/doi/10.1103/PhysRevX.7.011026}

\bibitem{pal2018}
Pal S, Nishad N, Mahesh T~S and Sreejith G~J 2018 {\em Phys. Rev. Lett.\/} {\bf
  120}(18) 180602
  \urlprefix\url{https://link.aps.org/doi/10.1103/PhysRevLett.120.180602}

\bibitem{iemini2018}
Iemini F, Russomanno A, Keeling J, Schir\`o M, Dalmonte M and Fazio R 2018 {\em
  Phys. Rev. Lett.\/} {\bf 121}(3) 035301
  \urlprefix\url{https://link.aps.org/doi/10.1103/PhysRevLett.121.035301}

\bibitem{rovny2018}
Rovny J, Blum R~L and Barrett S~E 2018 {\em Phys. Rev. Lett.\/} {\bf 120}(18)
  180603
  \urlprefix\url{https://link.aps.org/doi/10.1103/PhysRevLett.120.180603}

\bibitem{surace2019}
Surace F~M, Russomanno A, Dalmonte M, Silva A, Fazio R and Iemini F 2019 {\em
  Phys. Rev. B\/} {\bf 99}(10) 104303
  \urlprefix\url{https://link.aps.org/doi/10.1103/PhysRevB.99.104303}

\bibitem{wu2019}
Yu W~C, Tangpanitanon J, Glaetzle A~W, Jaksch D and Angelakis D~G 2019 {\em
  Phys. Rev. A\/} {\bf 99}(3) 033618
  \urlprefix\url{https://link.aps.org/doi/10.1103/PhysRevA.99.033618}

\bibitem{schaefer2019}
Sch\"afer R, Uhrig G~S and Stolze J 2019 {\em Phys. Rev. B\/} {\bf 100}(18)
  184301 \urlprefix\url{https://link.aps.org/doi/10.1103/PhysRevB.100.184301}

\bibitem{choi2019}
Choi J, Zhou H, Choi S, Landig R, Ho W~W, Isoya J, Jelezko F, Onoda S, Sumiya
  H, Abanin D~A and Lukin M~D 2019 {\em Phys. Rev. Lett.\/} {\bf 122}(4) 043603
  \urlprefix\url{https://link.aps.org/doi/10.1103/PhysRevLett.122.043603}

\bibitem{iadecola2019}
Iadecola T, Schecter M and Xu S 2019 {\em Phys. Rev. B\/} {\bf 100}(18) 184312
  \urlprefix\url{https://link.aps.org/doi/10.1103/PhysRevB.100.184312}

\bibitem{buca2019}
Bu{\v{c}}a B, Tindall J and Jaksch D 2019 {\em Nature Communications\/} {\bf
  10} 1730 ISSN 2041-1723
  \urlprefix\url{https://doi.org/10.1038/s41467-019-09757-y}

\bibitem{zhu2019}
Zhu B, Marino J, Yao N~Y, Lukin M~D and Demler E~A 2019 {\em New Journal of
  Physics\/} {\bf 21} 073028
  \urlprefix\url{https://doi.org/10.1088%2F1367-2630%2Fab2afe}

\bibitem{seibold2020}
Seibold K, Rota R and Savona V 2020 {\em Phys. Rev. A\/} {\bf 101}(3) 033839
  \urlprefix\url{https://link.aps.org/doi/10.1103/PhysRevA.101.033839}

\bibitem{pizzi2019}
Pizzi A, Knolle J and Nunnenkamp A 2019 {\em Phys. Rev. Lett.\/} {\bf 123}(15)
  150601
  \urlprefix\url{https://link.aps.org/doi/10.1103/PhysRevLett.123.150601}

\bibitem{matus2019}
Matus P and Sacha K 2019 {\em Phys. Rev. A\/} {\bf 99}(3) 033626
  \urlprefix\url{https://link.aps.org/doi/10.1103/PhysRevA.99.033626}

\bibitem{Heyl2020}
Russomanno A, Notarnicola S, Surace F~M, Fazio R, Dalmonte M and Heyl M 2020
  {\em Phys. Rev. Research\/} {\bf 2}(1) 012003
  \urlprefix\url{https://link.aps.org/doi/10.1103/PhysRevResearch.2.012003}

\bibitem{sacha2018}
Sacha K and Zakrzewski J 2017 {\em Reports on Progress in Physics\/} {\bf 81}
  016401 \urlprefix\url{https://doi.org/10.1088%2F1361-6633%2Faa8b38}

\bibitem{eckardt2017rmp}
Eckardt A 2017 {\em Rev. Mod. Phys.\/} {\bf 89}(1) 011004
  \urlprefix\url{https://link.aps.org/doi/10.1103/RevModPhys.89.011004}

\bibitem{rigol2014prx}
D'Alessio L and Rigol M 2014 {\em Phys. Rev. X\/} {\bf 4}(4) 041048
  \urlprefix\url{https://link.aps.org/doi/10.1103/PhysRevX.4.041048}

\bibitem{lazarides2014pre}
Lazarides A, Das A and Moessner R 2014 {\em Phys. Rev. E\/} {\bf 90}(1) 012110
  \urlprefix\url{https://link.aps.org/doi/10.1103/PhysRevE.90.012110}

\bibitem{rosch2015pra}
Genske M and Rosch A 2015 {\em Phys. Rev. A\/} {\bf 92}(6) 062108
  \urlprefix\url{https://link.aps.org/doi/10.1103/PhysRevA.92.062108}

\bibitem{russomanno2012prl}
Russomanno A, Silva A and Santoro G~E 2012 {\em Phys. Rev. Lett.\/} {\bf
  109}(25) 257201
  \urlprefix\url{https://link.aps.org/doi/10.1103/PhysRevLett.109.257201}

\bibitem{lazarides2014prl}
Lazarides A, Das A and Moessner R 2014 {\em Phys. Rev. Lett.\/} {\bf 112}(15)
  150401
  \urlprefix\url{https://link.aps.org/doi/10.1103/PhysRevLett.112.150401}

\bibitem{sen2019prb}
Maity S, Bhattacharya U, Dutta A and Sen D 2019 {\em Phys. Rev. B\/} {\bf
  99}(2) 020306
  \urlprefix\url{https://link.aps.org/doi/10.1103/PhysRevB.99.020306}

\bibitem{basko2006annp}
Basko D, Aleiner I and Altshuler B 2006 {\em Annals of Physics\/} {\bf 321}
  1126 -- 1205 ISSN 0003-4916
  \urlprefix\url{http://www.sciencedirect.com/science/article/pii/S0003491605002630}

\bibitem{huse2007prb}
Oganesyan V and Huse D~A 2007 {\em Phys. Rev. B\/} {\bf 75}(15) 155111
  \urlprefix\url{https://link.aps.org/doi/10.1103/PhysRevB.75.155111}

\bibitem{abanin2015prl}
Ponte P, Papi\ifmmode~\acute{c}\else \'{c}\fi{} Z, Huveneers F~m~c and Abanin
  D~A 2015 {\em Phys. Rev. Lett.\/} {\bf 114}(14) 140401
  \urlprefix\url{https://link.aps.org/doi/10.1103/PhysRevLett.114.140401}

\bibitem{abanin2016aop}
Abanin D~A, {De Roeck} W and Huveneers F 2016 {\em Annals of Physics\/} {\bf
  372} 1 -- 11 ISSN 0003-4916
  \urlprefix\url{http://www.sciencedirect.com/science/article/pii/S000349161630001X}

\bibitem{vishwanath2015naturec}
Bahri Y, Vosk R, Altman E and Vishwanath A 2015 {\em Nature Communications\/}
  {\bf 6} 7341 ISSN 2041-1723
  \urlprefix\url{https://doi.org/10.1038/ncomms8341}

\bibitem{potter2016prb}
Potter A~C and Vasseur R 2016 {\em Phys. Rev. B\/} {\bf 94}(22) 224206
  \urlprefix\url{https://link.aps.org/doi/10.1103/PhysRevB.94.224206}

\bibitem{else2016prb}
Else D~V and Nayak C 2016 {\em Phys. Rev. B\/} {\bf 93}(20) 201103
  \urlprefix\url{https://link.aps.org/doi/10.1103/PhysRevB.93.201103}

\bibitem{yao2017prl}
Yao N~Y, Potter A~C, Potirniche I~D and Vishwanath A 2017 {\em Phys. Rev.
  Lett.\/} {\bf 118}(3) 030401
  \urlprefix\url{https://link.aps.org/doi/10.1103/PhysRevLett.118.030401}

\bibitem{azbel1979prl}
Azbel M~Y 1979 {\em Phys. Rev. Lett.\/} {\bf 43}(26) 1954--1957
  \urlprefix\url{https://link.aps.org/doi/10.1103/PhysRevLett.43.1954}

\bibitem{andre1980aips}
Aubry S and Andr\'e G 1980 {\em Ann. Isr. Phys. Soc.\/} {\bf 3}(133)

\bibitem{thouless1983prb}
Thouless D~J 1983 {\em Phys. Rev. B\/} {\bf 28}(8) 4272--4276
  \urlprefix\url{https://link.aps.org/doi/10.1103/PhysRevB.28.4272}

\bibitem{wilkinson1984prsa}
Wilkinson M 1984 {\em Proc. R. Soc. Lond. A\/} {\bf 391} 305--350
  \urlprefix\url{https://royalsocietypublishing.org/doi/10.1098/rspa.1984.0016}

\bibitem{sokoloff1985physrep}
Sokoloff J 1985 {\em Physics Reports\/} {\bf 126} 189 -- 244 ISSN 0370-1573
  \urlprefix\url{http://www.sciencedirect.com/science/article/pii/0370157385900882}

\bibitem{kohmoto1989prl}
Hiramoto H and Kohmoto M 1989 {\em Phys. Rev. Lett.\/} {\bf 62}(23) 2714--2717
  \urlprefix\url{https://link.aps.org/doi/10.1103/PhysRevLett.62.2714}

\bibitem{igloi1993jpa}
Igloi F 1993 {\em Journal of Physics A: Mathematical and General\/} {\bf 26}
  L703--L709
  \urlprefix\url{https://doi.org/10.1088%2F0305-4470%2F26%2F15%2F016}

\bibitem{laumann2017prx}
Chandran A and Laumann C~R 2017 {\em Phys. Rev. X\/} {\bf 7}(3) 031061
  \urlprefix\url{https://link.aps.org/doi/10.1103/PhysRevX.7.031061}

\bibitem{note1}
  However, on the phase boundarie $\varphi_J$ could have important influence on
  the scaling of the localization length of low-energy excitations, as gap
  vanishes there and the functional form of the localization length should
  connects smoothly to that of the edge modes
  \cite{laumann2018prl,laumann2018arxiv}

\bibitem{laumann2018prl}
Crowley P~J~D, Chandran A and Laumann C~R 2018 {\em Phys. Rev. Lett.\/} {\bf
  120}(17) 175702
  \urlprefix\url{https://link.aps.org/doi/10.1103/PhysRevLett.120.175702}

\bibitem{laumann2018arxiv}
Crowley P~J~D, Chandran A and Laumann C~R 2018 Critical behavior of the
  quasi-periodic quantum ising chain (\textit{Preprint} \eprint{1812.01660})

\bibitem{sen1996}
Chakrabarti B~K, Dutta A and P S 1996 {\em Transverse Ising Chain (Pure
  System)\/} (Springer)

\bibitem{sondhi2016prb}
von Keyserlingk C~W and Sondhi S~L 2016 {\em Phys. Rev. B\/} {\bf 93}(24)
  245146 \urlprefix\url{https://link.aps.org/doi/10.1103/PhysRevB.93.245146}

\bibitem{santoro2012prl}
Ziraldo S, Silva A and Santoro G~E 2012 {\em Phys. Rev. Lett.\/} {\bf 109}(24)
  247205
  \urlprefix\url{https://link.aps.org/doi/10.1103/PhysRevLett.109.247205}

\bibitem{santoro2013prb}
Ziraldo S and Santoro G~E 2013 {\em Phys. Rev. B\/} {\bf 87}(6) 064201
  \urlprefix\url{https://link.aps.org/doi/10.1103/PhysRevB.87.064201}

\end{thebibliography}

\end{document}